\documentclass[aps,prd,twocolumn,showpacs,preprintnumbers,nofootinbib,amsmath,amssymb,floatfix]{revtex4}

\usepackage{graphicx}
\usepackage{dcolumn}
\usepackage{bm}
\usepackage{amsmath}
\usepackage[normalem]{ulem}

\newcommand{\ie}{\emph{i.e.}\,\,}
\newcommand{\eg}{\emph{e.g.}}

\newcommand{\tr}{\ensuremath{\mathrm{Tr}}}
\newcommand{\re}{\ensuremath{\mathrm{Re}}}
\newcommand{\im}{\ensuremath{\mathrm{Im}}}

\DeclareRobustCommand{\Eqref}[1]{Eq.~(\ref{#1})}
\DeclareRobustCommand{\Figref}[1]{Fig.~\ref{#1}}

\begin{document}

\title{Comparison of the gradient flow with cooling in $SU(3)$ pure gauge theory}

\author{Claudio Bonati}
\email{bonati@df.unipi.it}
\affiliation{
Dipartimento di Fisica dell'Universit\`a
di Pisa and INFN - Sezione di Pisa,\\ Largo Pontecorvo 3, I-56127 Pisa, Italy}

\author{Massimo D'Elia}
\email{delia@df.unipi.it}
\affiliation{
Dipartimento di Fisica dell'Universit\`a
di Pisa and INFN - Sezione di Pisa,\\ Largo Pontecorvo 3, I-56127 Pisa, Italy}

\begin{abstract}
The gradient (Wilson) flow has been introduced recently in order to 
provide a solid theoretical framework for the 
smoothing of ultraviolet noise in lattice gauge configurations. 
It is interesting to ask 
how it compares with other, more heuristic and numerically cheaper 
smoothing techniques, such as standard cooling.
In this study we perform
such a comparison, focusing on observables related to topology.
We show that, already for moderately small lattice spacings, 
standard cooling and the gradient flow lead to
equivalent results, both for average quantities and configuration by 
configuration.
\end{abstract}

\maketitle

\section{Introduction}

At present, the lattice formulation represents the best 
available tool for a gauge-invariant regularization~\cite{Wilson} 
and a systematic
non-perturbative numerical study 
of strong interactions, and, more generally, of gauge theories.
Like for any regularized theory, one has to deal with unphysical
fluctuations at the scale of the ultraviolet (UV) cutoff (which is 
set by the lattice spacing in the case of lattice gauge theories),
which must be properly treated. 
Over the years, various techniques have been developed in this sense.
Where possible, proper prescriptions can be assigned for a perturbative
or non-perturbative renormalization of physical quantities. Another 
widely used technique is instead to apply some kind of smoothing procedure,
in order to dampen the fluctuations at the UV scale,
while hopefully leaving the physical content unchanged.

Typical examples of observables requiring renormalization are 
the topological charge (winding number) and, more generally, 
the observables related to topology, like the topological susceptibility. 
Contrary to its continuum counterpart, the lattice gluonic definition of the 
topological charge is affected by UV fluctuations: 
it gets multiplicatively renormalized~\cite{zetaref}
and does 
not take integer values. Further additive renormalizations, 
related to contact terms, 
appear when defining the topological susceptibility~\cite{addref}.

Apart from switching to a fermionic definition of 
topology, via the index theorem, 
various strategies have been developed
to successfully deal with such renormalizations, going from
a direct computation and subtraction of 
them~\cite{teper-89,cdpv-90,dv-92,acdgv-93,fp-94,add-97,addk-98} 
to the 
application of smoothing methods to dampen the UV fluctuations
and recover an almost integer valued observable.
In this respect, cooling techniques~\cite{cooling}, which proceed through a local minimization
of the gluonic action, are particularly well suited, since the topological 
content of gauge configurations becomes quasi-stable against minimization
as one approaches the continuum limit, \ie as one recovers
a proper definition of the gauge field topology.

In this context, the recent introduction of the gradient flow (also known as 
Wilson flow when used in connection with the Wilson action) represents an 
important advance~\cite{Luscher_wf0, Luscher_wf1}. 
The main difference with respect to previously used smoothers 
is that in this case the elimination of UV fluctuations 
is governed by a differential equation, 
thus achieving a better analytical control of the smoothing procedure.

An interesting and due question regards how the gradient flow compares 
with other standard smoothing techniques.
The present study is a step in the direction of clarifying 
this issue.
In particular we will compare standard cooling and the gradient flow, 
for the determination of topological observables,
in the $SU(3)$ lattice gauge theories discretized with the Wilson action.

Such a question is of general interest, since standard 
smoothing techniques have been widely used in the literature. 
But it is also of great practical interest.
Indeed, as we will clarify later, the application
of the gradient flow is much more computationally demanding 
than standard cooling.
It is therefore compelling to 
understand what the effective differences are between the two methods, 
depending on the chosen observable.

The paper is organized as follows. In Section~\ref{sec2}
we provide a brief description of the gradient flow 
and of standard cooling and then compare them
in the limit of smooth fields. In
Section~\ref{sec3} we first discuss 
the definition of a general setting for the comparison of the two methods;
then we present our numerical results for
topological quantities.
Finally, in Section~\ref{sec4}, we discuss our results
and draw our conclusions.

\section{Cooling and the gradient flow}
\label{sec2}

In this section, for the benefit of the reader, we recall the definitions of the 
gradient flow and of the cooling procedure, and present the details of our implementation. 
We will limit ourselves to the case of the Wilson action
for $SU(N)$ pure gauge theories~\cite{Wilson}, which is the 
one used in our simulations, but the generalization to different
discretizations does not present significant difficulties.

The Wilson action is written in terms of the product of the link variables $U_{\mu}(x)$ along an 
elementary face (plaquette) of the lattice, $U_{\mu\nu}(x)$, in the form
\begin{equation}\label{wilson_act}
S=\frac{2N}{g_0^2} \sum_{\substack{x,\, \mu < \nu \\ \mu,\nu\ge 0}}\left(1-\frac{1}{N}\re\tr U_{\mu\nu}\right)\ ,
\end{equation} 
where $N$ is the number of colors.
It is convenient to introduce the staples as the (in general non unitary) matrices $W_{\mu}(x)$, defined by
\begin{equation}\label{staples}
\begin{aligned}
W_{\mu}(x)=&\sum_{\nu\ge 0, \nu\neq\mu}\Big[ U_{\nu}(x)U_{\mu}(x+\hat{\nu})U_{\nu}^{\dag}(x+\hat{\mu})+\\
&+U_{\nu}^{\dag}(x-\hat{\nu})U_{\mu}(x-\hat{\nu})U_{\mu}(x-\hat{\nu}+\hat{\mu})\Big]\ ;
\end{aligned}
\end{equation}
the part of the action involving a given link variable $U_\mu(x)$ is then
simply written as $-(2/g_0^2) \re\tr [U_\mu(x) W_{\mu}^{\dag}(x)]$.
To avoid confusion, 
in the following expressions we will not make use of the implicit 
summation over repeated indices.

\subsection{Gradient flow}

The gradient flow is defined (see \cite{Luscher_wf0, Luscher_wf1}) by the solution of the evolution equations
\begin{equation}\label{wf}
\begin{aligned}
&\dot{V}_{\mu}(x,\tau)=- g_0^2\big[\partial_{x,\mu}S(V(\tau))\big]V_{\mu}(x,\tau) \\
& V_{\mu}(x,0)=U_{\mu}(x)\ ,
\end{aligned}
\end{equation}
where the link derivatives are defined by
\begin{equation}
\begin{aligned}
\partial_{x,\mu} f(U)&=i\sum_a T^a \left.\frac{\mathrm{d}}{\mathrm{d}s}f(e^{isX^a}U)\right|_{s=0} \\
&\equiv i \sum_a T^a \partial_{x,\mu}^{(a)} f(U)\ .
\end{aligned}
\end{equation}
In this expression $T^a$ are the (hermitian)~\footnote{
We would like to warn the reader that such a notation
is different from the one adopted in the literature
where the gradient flow has been originally 
discussed~\cite{Luscher_wf0, Luscher_wf1}.
We follow here the standard convention
in which the generators of $SU(N)$ are taken to be hermitian,
\eg, for $SU(3)$, $T^a = \lambda^a/2$ where $\lambda^a$ are
the Gell-Mann matrices.}
generators of the $SU(N)$ algebra, 
with the normalization $\tr(T^aT^b)=\frac{1}{2}\delta^{ab}$, and 
\begin{equation}
X^a(y,\nu)=\left\{\begin{array}{ll} T^a & \mathrm{if}\ (y,\nu)=(x,\mu) \\ 0 & 
\mathrm{else} \end{array} \right. \ .
\end{equation}

If we introduce the notation $\Omega_{\mu}=U_{\mu}(x)W_{\mu}^{\dag}(x)$ we have
\begin{equation}
\partial_{x,\mu}^{(a)}S(U)=\frac{2}{g_0^2}\im\tr\big[T^a \Omega_{\mu}\big]
\end{equation}
and $\partial_{x,\mu}S(U)$ is given by
\begin{equation}\label{generator}
\begin{aligned}
&g_0^2\, \partial_{x,\mu}S(U)=2i\sum_a T^a\im\tr\big[T^a \Omega_{\mu}\big]=\\
&=\frac{1}{2}\left(\Omega_{\mu}-\Omega^{\dag}_{\mu}\right)-
\frac{1}{2N}\tr\left(\Omega_{\mu}-\Omega^{\dag}_{\mu}\right)\ .
\end{aligned}
\end{equation}

In practice, the gradient flow moves the gauge 
configuration along the steepest descent direction in the configuration
space, \ie along the gradient of the action (hence the 
name of gradient flow); in particular, the chosen sign in the evolution
equations leads to a minimization of the action. Indeed,
from the definition \Eqref{wf} and the previous expressions it is simple to show that
\begin{equation}
\frac{\mathrm{d}}{\mathrm{d}\tau}S(V(\tau))=-g_0^2\,\sum_{a, x,\mu\ge 0} 
\big[\partial_{x,\mu}^{(a)} S(V(\tau))\big]^2 \le 0\ .
\end{equation} 
Thus $S(V(\tau))$ is a monotonically decreasing function of the flow time $\tau$ and 
the ``evolved'' variables $V_{\mu}(x,\tau)$ can be used as the smoothed version of the original
link variables $U_{\mu}(x)$. From the explicit expression, 
\Eqref{generator}, we can also see that the 
gradient flow is just the flow generated by the infinitesimal version of the 
isotropic stout smearing introduced in Ref.~\cite{MP}.
It is important to stress at this point that quantity $\tau$, 
defined here and used in the following,
is the flow time in dimensionless units.

The integration of the flow in \Eqref{wf} can be performed by using standard methods for 
ordinary differential equations; in particular, we adopted the third order
Runge-Kutta scheme described in appendix C 
of Ref.~\cite{Luscher_wf1}. We have chosen an elementary integration step $\epsilon=0.02$, and verified that 
the integration error induced by this choice does not significantly affect our results~\footnote{In particular, 
results obtained on a subsample of configurations by using a
different integration step, $\epsilon=0.01$, are indistinguishable 
from those obtained at $\epsilon=0.02$.}.

\subsection{Cooling}

Also in the case of cooling, 
the idea is that of evolving the gauge configuration
so as to minimize the gauge action: in fact, cooling has been one 
of the first procedures introduced to get rid of UV artifacts
by smoothing gauge configurations~\cite{cooling}.
However, while the gradient flow is defined by an evolution equation, 
the cooling method proceeds by discrete steps: in each step
the action is minimized with respect to a subset of configuration
variables (\eg, a single link or even a link subgroup), and then the 
procedure is performed iteratively over all variables,
in order to achieve a global movement of the configuration
towards the minimum of the gauge action.

Many variants of cooling have been devised, in which 
the discrete steps are made more or less smooth~\cite{coolcon,coolnuc}. 
Here we will
consider the simplest, original version, also known as 
standard cooling:
\begin{itemize} 
\item An elementary step of the algorithm consists in 
replacing a given link variable  
$U_{\mu}(x)$ by the group element which locally minimizes
the action, while all other link variables are kept
fixed, \ie by the matrix $M\in SU(N)$ which maximizes
\begin{equation*}
\re\tr\Big[ M W_{\mu}^{\dag}(i)\Big] \, .
\end{equation*}
In the particular case of the $SU(2)$ gauge theory,
the maximization step can be performed analytically, with the result
\begin{equation*}
M=\frac{W_{\mu}(x)}{\sqrt{\det W_{\mu}(x)}}\ .
\end{equation*}
For $SU(N)$ the maximization is performed by using a Cabibbo-Marinari like algorithm (\cite{CM}),
\ie by iterating the maximization over a covering set of $SU(2)$ subgroups.\\
\item The procedure is then repeated iteratively by visiting
link variables on all sites and along all directions of the lattice.
In our implementation we will first sweep lattice sites, following the standard
lexicographic order, and then link directions, starting with 
$x = (0,0,0,0)$ and $\mu=0$. Both the starting link and the visiting 
order can be changed at will, 
leading to slightly different cooling variants.\\
\item A complete sweep of the lattice is what is usually
called a \emph{cooling step}. 
A cooling step can be iterated $n_c$ times, 
thus generating a (discrete) flow in the space of gauge configurations. 
\end{itemize}

It is important to stress once more that, contrary to other smoothing
procedures such as smearing, a cooling sweep proceeds iteratively, \ie
at each elementary step the cooled link is substituted in the 
configuration before computing the staple needed to cool the next link. 
If staples were all computed before starting the cooling sweep, the 
decrease of the action would 
not be guaranteed anymore, and instabilities would appear, similar
to those happening in the repeated application of smearing when
the smearing parameter is too large (see, \eg, Ref.~\cite{jim}). 
This iterative nature of cooling
will be important when discussing the speed at which cooling
proceeds, as compared with the gradient flow.

It is interesting to notice that there is one variant of cooling
which resembles the gradient flow more closely, namely the 
controlled cooling
introduced in Ref.~\cite{coolcon}. In that case, the elementary step
of cooling consists in minimizing the action under the constraint
\begin{equation} \label{controlled}
\frac{1}{N} \tr \big\{ (U_\mu^\dagger - U_\mu'^\dagger) (U_\mu - U_\mu')  \big\}
\leq \delta^2
\end{equation}
where $U_\mu$ and $U_\mu'$ are, respectively, the old and the new 
link variables. Also in this case the configuration proceeds towards
a minimum, but with the constraint that at each elementary 
step the new link variable
does not differ much from the old variable, depending on the value
of the controlling parameter $\delta$. For small enough $\delta$,
the cooling step effectively becomes an infinitesimal movement along the steepest
descent direction, \ie it becomes a possible integrator of the 
gradient flow. Indeed, the authors of Ref.~\cite{coolcon} 
verified that, for small enough $\delta$, the order in which
the links are cooled becomes immaterial. 

\begin{table}
\begin{tabular}{|c|c|c|}
\hline \rule{0mm}{3.5mm}$\beta=6/g_0^2$ & $r_0$ & $a$ \\ \hline
5.95 &  4.898(12) & 0.1021(25) \\ \hline 
6.07 &  6.033(17) & 0.08288(23) \\ \hline
6.2  &  7.380(26) & 0.06775(24) \\ \hline
\end{tabular}
\caption{\label{table1}
Values of the bare couplings used in this work, of the corresponding 
Sommer scale (computed in Ref.~\cite{GSW}) and of the lattice spacing, evaluated by 
using the reference value $r_0=0.5\,\mathrm{fm}$.}
\end{table}

\subsection{Perturbative relation between the two smoothing procedures}

As we have already stressed, both cooling and the gradient flow
evolve the gauge configuration towards a minimum of the 
gauge action. In a perturbative approximation, in which 
all link variables are very close to the identity element
of the gauge group, the connection between the two procedures
can be investigated in more detail, and a relation can be found
between the speed at which the two evolutions proceed. This relation
will be compared with the numerical results in 
Section~\ref{sec3}.

Let us assume, therefore, that $U_\mu(x) \simeq 1 + i\,\sum_a u_\mu^a(x) T^a$ 
for each link variable, so that the staple takes the simple 
form $W_\mu(x) \simeq 6 + i\,\sum_a w_\mu^a(x) T^a$, where
both $u_\mu^a(x)$ and  $w_\mu^a(x)$ are infinitesimal quantities.
In this approximation, one has 
\begin{equation*}
\Omega_\mu \simeq 6 + i\,\sum_a \big[6 u_\mu^a(x) - w_\mu^a(x)\big] T^a
\end{equation*}
and Eq.~(\ref{generator}) becomes
\begin{equation}\label{generatorpert}
g_0^2\, \partial_{x,\mu}S(U)=
i\,\sum_a \big[6 u_\mu^a(x) - w_\mu^a(x)\big] T^a\, .
\end{equation}
As a consequence, the 
evolution equation of the gradient
flow can be approximated as follows:
\begin{equation}\label{pert_wflow}
u_\mu^a (x,\tau + \epsilon) \simeq 
u_\mu^a (x,\tau) - \epsilon \big[6 u_\mu^a(x,\tau) - w_\mu^a(x,\tau)\big] \, .
\end{equation}

On the other hand, cooling acts so as to substitute $U_\mu (x)$ with
the projection of $W_\mu(x)$ 
over the gauge group. In the perturbative
approximation, this  projection is simply 
$1 + i\,\sum_a (w_\mu^a(x)/6)\, T^a$, so that the elementary 
cooling step corresponds to the substitution
\begin{equation}\label{pert_cooling}
u_\mu^a (x) \to \frac{w_\mu^a(x)}{6} \, .
\end{equation}

A naive comparison of Eqs.~(\ref{pert_wflow}) and (\ref{pert_cooling})
would lead to the conclusion that the instantaneous speed at which links
evolve in the gradient flow is such that a whole cooling step would
be covered in a step $\epsilon = 1/6$ of gradient flow evolution, \ie
that the approximate relation $\tau \simeq n_c/6$ should
hold between the gradient flow time and the number of cooling 
steps. The factor 6 comes, given the normalization chosen
for the gradient, from the number of staples around a given 
link, \ie it is equal to $2\, (D - 1)$, where $D$ is the number 
of space-time dimensions.

However, such a conclusion is wrong by a factor $2$, as is clear
from the following argument. The staple appearing in the 
gradient flow is constructed with gauge links all computed at the 
same flow time $\tau$. On the contrary, in the case of cooling, due
to the iterative nature of the process, some of the links used
to construct the staple 
have already undergone the cooling step under consideration,
and this results in an increase in the speed of cooling.
For a regular 
visiting order of the lattice links during the sweep,
one has that, on average, that half of the neighboring links have already
been cooled one more time: that results in a speed increase for cooling by a factor 2 with 
respect to the naive expectation,
as one can evince
from the simple diffusive model discussed
in Appendix~\ref{appendA}.

Therefore, the
predicted perturbative relation is actually 
$\tau \simeq n_c / (D - 1) = n_c / 3$. Such a relation is expected
to depend on the dimensionality of the system (and 
on the normalization of the gradient, \ie on the fact that 
we actually take the gradient of $g_0^2 S$), but not
on the number of colors, at least in the limit of smooth
fields.

\subsection{Smoothing and the continuum limit}

Let us now discuss how smoothing has to be 
tuned as the continuum limit is approached. An important point is that
this tuning is independent of the particular kind of smoothing,
be it cooling, the gradient flow or something else, 
once a precise correspondence 
has been found between the different techniques, 
which is valid lattice spacing by lattice spacing.

Smoothing is, in general, an arbitrary modification of the theory in the UV, 
up to some length scale 
$\lambda_{S}$, with the only requirement 
that it dampens the quantum fluctuations
on length scales smaller than $\lambda_S$. 
While smoothing changes the theory up to $\lambda_S$, we have to ensure that 
this does not affect our continuum results, \ie that physics does not depend on 
the choice of $\lambda_S$.
If we are studying an observable which is naturally defined for large distances only, an 
obvious example being the effective mass extracted from the expectation value of a correlator,
then it is natural to keep $\lambda_S$ fixed in physical units:  to avoid systematical dependences on 
$\lambda_S$, it will be sufficient to use correlators defined at distances $r\gg \lambda_S$. 

This possibility is particularly appealing in the gradient flow 
setup, since it can be shown that composite operators defined at fixed physical flow time 
renormalize in a simple way (see Ref.~\cite{Luscher:2011bx}).
In particular, for the case of the gradient flow one has
\begin{equation}\label{lambda_wilson}
\lambda_S\simeq \sqrt{8t}\ ,
\end{equation}
where $t = a^2 \tau$ is the flow time in physical units 
(see Ref.~\cite{Luscher_wf1}),
$a$ being the lattice spacing. This procedure can now be simply
translated in terms of cooling. Indeed
from the argument of the 
previous section (\ie $\tau\simeq n_c/3$), which will
be accurately verified against numerical results in the following sections, 
we expect for cooling the analogous relation
\begin{equation}\label{lambda_cooling}
\lambda_S\simeq a\sqrt{8n_c/3}\ ,
\end{equation}
i.e. the number of cooling steps has to be scaled proportionally to 
$1/a^2$ in order to keep $\lambda_S$ fixed. Actually
this is not a completely new result, since it is already well known that
cooling acts like a diffusive process.

The situation can be less trivial
for observables which are not related to large distance correlators,
but are instead an integral over all distances of some two 
point function, like a susceptibility.
In this case it is not guaranteed apriori that keeping $\lambda_S$ 
fixed will not affect the continuum limit, and one must look for 
the existence of a proper ``safe scaling window'' for $\lambda_S$
(see Ref.~\cite{Luscher:2013vga} for a discussion regarding the 
gradient flow). 

An example is the topological susceptibility, which is the integral over all
distances of the two point correlator of the topological charge density.
In this case one can follow different strategies to look for
the safe scaling window. 
A known procedure~\cite{vicari_rep} 
is to look for a plateau in terms of $\lambda_S$ at every fixed lattice spacing,
and then perform the continuum extrapolation of the plateau values.
The existence of the plateau ensures that $\lambda_S$ is small enough not to affect the physical result and 
that, on the other hand, the smoothing is effective in removing additive and multiplicative renormalizations.

Alternatively, one could perform the continuum limit of results obtained at fixed $\lambda_S$, and 
then look for a safe plateau, in  terms of $\lambda_S$, in the continuum extrapolated values.
It is not the purpose of this study to perform an accurate check
of the consistency of these two strategies.
What we will show, instead,  is that two perfectly equivalent definitions of 
$\lambda_S$ exist at every fixed lattice spacing, \Eqref{lambda_wilson} and \Eqref{lambda_cooling},
defined by either cooling or the gradient flow, in terms of which one can 
perform the preferred continuum extrapolation.

\section{Numerical results}
\label{sec3}

Most of the simulations have been performed on a $20^4$ lattice at the bare 
coupling values $6/g_0^2=5.96, 6.07, 6.2$, corresponding to the lattice spacings 
reported in Table~\ref{table1} and to physical lattice sizes ranging from 
$2\, \mathrm{fm}$ to $1.4\, \mathrm{fm}$.
Here We do not have the aim to keep finite size effects well under control,
since our purpose is simply the check how cooling and the gradient flow
compare to each other, on the same configuration sample, in the smoothing of
fluctuations at the UV scale. However, 
a comparison with some simulations performed on larger lattices
shows that such effects are not large and do not significantly affect  
our conclusions.

For each value of the bare coupling we have generated $\mathcal{O}(10^4)$ 
configurations, 
each one separated from the next by $200$ Monte Carlo steps, a single step 
consisting of a full lattice update with $1$ heatbath (\cite{Creutz1980, KP}) and 
$5$ overrelaxation sweeps (\cite{Creutz1987}). 
On these configurations, we have evaluated 
the topological charge after smoothing,
by using both cooling (we have reached a  maximum of $50$ cooling steps, 
with measurements taken after each
step) and the gradient flow (reaching a maximum flow time $\tau=10$,
with measurements performed
every $\Delta\tau=0.2$). The expression used for the discretization of the 
topological charge density is 
\begin{equation}\label{qL}
q_L(x) = -\frac{1}{2^9 \pi^2}\sum_{\mu\nu\rho\sigma = \pm 1}^{\pm 4} 
\tilde{\epsilon}_{\mu\nu\rho\sigma} \tr \left(U_{\mu\nu}(x) U_{\rho\sigma}(x) \right) \; ,
\end{equation}
where ${\tilde{\epsilon}}_{\mu\nu\rho\sigma} = \epsilon_{\mu\nu\rho\sigma}$ for positive 
indices, while for the negative directions the relation 
${\tilde{\epsilon}}_{\mu\nu\rho\sigma} = -{\tilde{\epsilon}}_{(-\mu)\nu\rho\sigma}$ 
and the complete antisymmetry are used.

\subsection{Setting a common scale}

\begin{figure}[t!]
\includegraphics[width=0.9\columnwidth, clip]{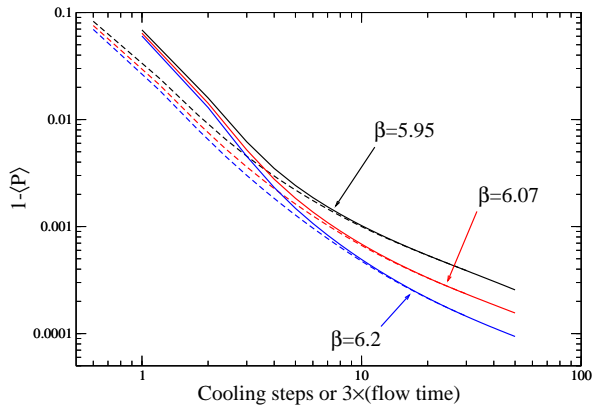}
\caption{Behavior of one minus the average plaquette as a function of number of cooling steps (continuous lines) and of 
(three times) the gradient flow time (dashed lines).} 
\label{step_time}
\end{figure}

The purpose of the present study is to compare how the (continuous) gradient
flow and the discrete flow generated by cooling compare to each other.
It is clear that, in order to do that, we need to set a common scale,
\ie to fix \emph{apriori} what is the flow time $\tau$ 
to be compared with $n_c$ 
cooling steps. 
The simplest way to proceed is to set such a common scale 
by using some standard observable, and the most natural observable is 
given by the quantity whose minimization defines both flows, that is by the
action itself.
This is also the strategy adopted in the past to compare 
different versions of cooling~\cite{coolcomp}.

In \Figref{step_time} we report 
the average plaquette (action density) 
values as a function of $n_c$ (for cooling) and
of $3\tau$ (for the gradient flow case). Such functions
permit to obtain the desired correspondence:
for each given value of the inverse bare gauge coupling $\beta$, 
we define $\tau(n_c)$ as the value of the gradient time $\tau$ 
that changes the average action 
by the same amount as $n_c$ cooling steps. 

\begin{figure}[t!]
\includegraphics[width=0.9\columnwidth, clip]{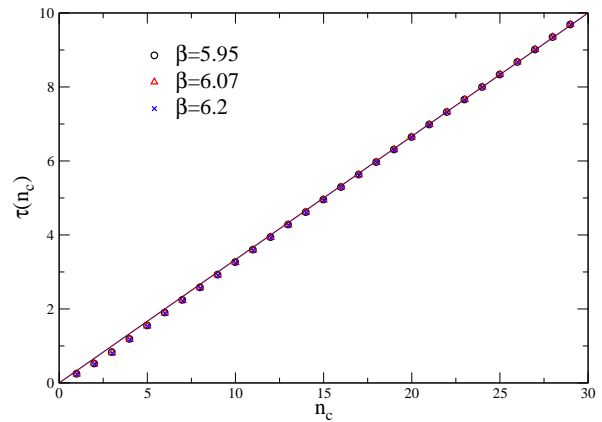}
\caption{Behavior of $\tau(n_c)$ as a function of the number of cooling steps $n_c$. The continuous line 
corresponds to  $\tau=n_c/3$. Data points at different lattice spacings are hardly distinguishable.}
\label{step_time3}
\end{figure}

A plot of the functions $\tau(n_c)$,
obtained for the different explored values
of $\beta$, is shown in \Figref{step_time3}:
the agreement among the different lattice spacings
is striking and demonstrates that the correspondence between cooling and the 
gradient flow has a perfectly well-defined continuum limit. 
The continuous line corresponds to the function
$\tau=n_c/3$. It is clear that this function is
a good approximation of $\tau(n_c)$ for all the lattice spacings used, 
and becomes better and better as $n_c$ increases: the agreement is 
at the level of 1\%
for $n_c = 10$ and of 0.1\% (\ie already within the precision
of our determination) for $n_c = 20$.
In the following, for simplicity, 
we will just use the approximation $\tau(n_c)\simeq n_c/3$, which is equivalent to saying that
one unit of gradient flow time corresponds to three cooling steps; corrections
to this assumptions prove to be completely irrelevant to the following
analysis. {Preliminary results show that the relation 
$\tau(n_c)\simeq n_c/3$ holds true also for the gauge group $SU(2)$,
thus supporting the perturbative argument of the previous section.
}

\subsection{Determination of the topological background}

The lattice topological charge $Q_L$ is defined as the sum over all the lattice sites of the
charge density given in \Eqref{qL}. 
Although $Q_L$ is not exactly quantized, because of lattice artifacts, 
sharp peaks appear in the topological charge distribution, 
as the smoothing procedure goes on,
located at approximately integer values.

\begin{figure}[t!]
\includegraphics[width=0.9\columnwidth, clip]{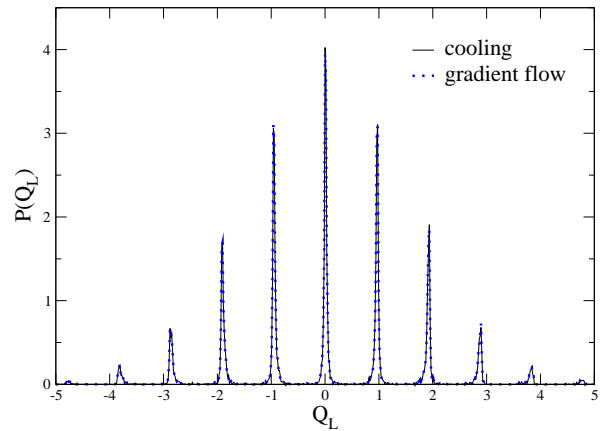}
\caption{Probability distribution of the topological charge for $\beta=6.2$,
evaluated after $21$ cooling steps and after gradient flow with $\tau=7$.
Due to the very good agreement, 
the two distributions are hardly distinguishable in the figure.
The corresponding figures for the other $\beta$ values are analogous.
}
\label{qdistr}
\end{figure}

An example of the probability distribution $P(Q_L)$ of $Q_L$ for $\beta=6.2$ is shown in \Figref{qdistr}, 
where the results obtained both with cooling ($n_c=21$) and the gradient flow ($\tau=7$) are shown.
The fact that the two distributions perfectly agree with each other is a first indication that,
at least for the computation of average quantities, 
the two considered smoothing procedures are equivalent.

In order to reduce the lattice artefacts and improve the convergence 
towards the continuum limit,
the estimator $Q$ of the topological background 
that will be used in the following analysis 
is defined by the procedure \cite{b4-1}:
\begin{equation}\label{Qdef}
Q=\mathrm{round}\left(\alpha\, Q_L \right)\ ,
\end{equation} 
where $\mathrm{round}(x)$ denotes the integer closest to $x$ and the rescaling factor $\alpha$ is determined 
in such a way to minimize
\begin{equation}\label{tomin}
\left\langle \left( \alpha\, Q_L - \mathrm{round}
\left[\alpha\, Q_L \right]\right)^2\right\rangle \ .
\end{equation}
In this way, the distribution of $\alpha Q_L$ is such that
the sharp peaks visible in \Figref{qdistr} move exactly onto integer 
values. 

We emphasize that this procedure is not a renormalization,
but just a redefinition of the observable in order to obtain an integer-valued topological
charge and to significantly reduce lattice artefacts, see \cite{vicari_rep} for
a discussion on this point. On the other hand,
as we will show in the following (see Table~\ref{table2}),
cooling and the gradient flow lead to perfectly equivalent results
independently of the chosen definition of topological charge. This is 
also manifest from Fig.~\ref{qdistr}, where no rounding has been applied.

\begin{figure}[t!]
\includegraphics[width=0.9\columnwidth, clip]{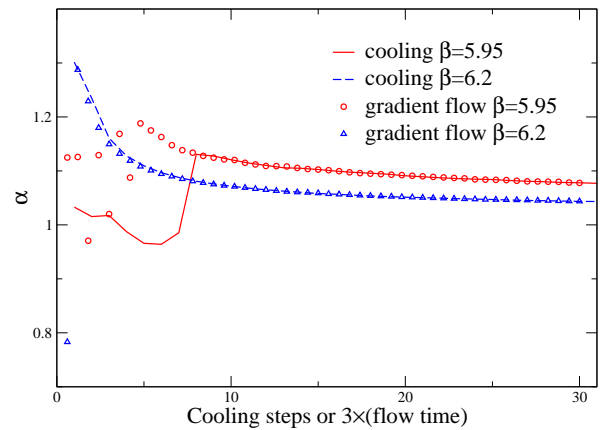}
\caption{Plot of the rescaling factor $\alpha$ to be used in the definition of the
topological charge, \Eqref{Qdef}, for $\beta=5.95$ and $\beta=6.2$, when using cooling 
and gradient flow.
}
\label{rescaling}
\end{figure}

An example of the behaviour of the rescaling factor $\alpha$ is reported in \Figref{rescaling}, 
for two different values of $\beta$.
The oscillations observed for a small number of cooling steps (or equivalently for small values of 
the flow time) are due to instabilities of the optimization procedure adopted to minimize 
\Eqref{tomin} and disappear once the configurations are smooth enough (\ie once the peaks in 
$P(Q_L)$ are well defined); in particular, they almost disappear by reducing the lattice spacing.

\subsection{Comparison for average quantities: the topological susceptibility}

In this section we present our results for the topological susceptibility obtained by using 
cooling or the gradient flow as smoothers. The topological susceptibility is defined by
\begin{equation}\label{chidef}
\chi=\frac{\langle Q^2\rangle }{V} =
\frac{\langle Q^2\rangle }{a^4 N_tN_s^3}\ ,
\end{equation}
where $N_t$ and $N_s$ are the temporal and spatial extents of the lattice and $Q$ is
given by \Eqref{Qdef}. 

In the continuum limit, topological sectors become strictly 
separated and the topological charge $Q$ is stable under 
any smoothing procedure which minimizes the action. 
On the other hand, 
at finite lattice spacing, $Q$ is in general
only quasi-stable, 
and topological backgrounds can be eventually washed out by 
a prolongated smoothing. However
the two time scales, at which the UV fluctuations or the topological background
are respectively affected, become rapidly well separated as the 
lattice spacing is reduced. That results in the appearance 
of a well-defined and extended
plateau, as a function of the amount of smoothing, 
for topological quantities, like the susceptibility $\chi$.

\begin{figure}[t!]
\includegraphics[width=0.9\columnwidth, clip]{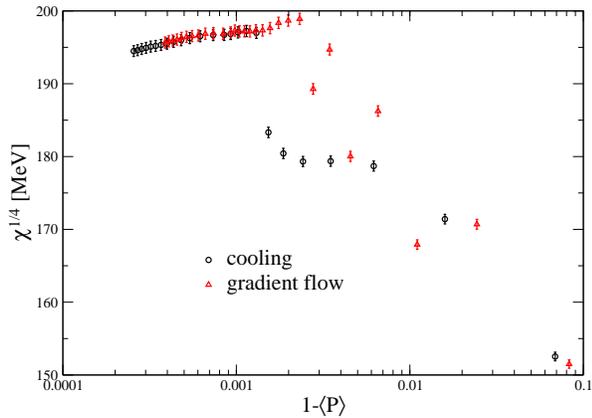}
\caption{Topological susceptibility computed for $\beta=5.95$ after cooling or gradient flow. The average plaquette 
$\langle P\rangle$ is used to set a common scale.}
\label{chi_5.95}
\end{figure}

\begin{figure}[t!]
\includegraphics[width=0.9\columnwidth, clip]{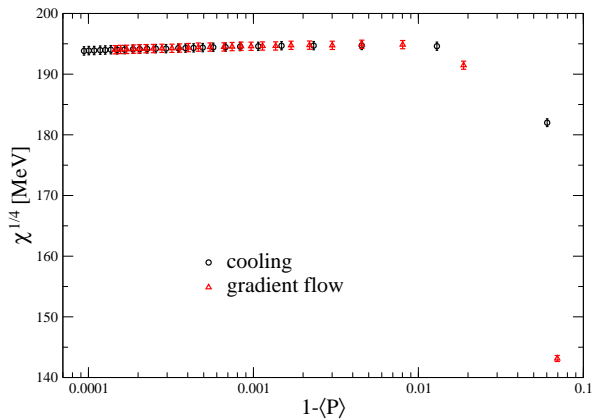}
\caption{Topological susceptibility computed for $\beta=6.2$ after cooling or gradient flow. The average plaquette 
$\langle P\rangle$ is used to set a common scale.}
\label{chi_6.2}
\end{figure}

All this is well known for cooling
and, for the reason previously explained, the plateau value
is the one typically used in computations.
Our purpose is now to check if, under the gradient flow, the topological 
susceptibility behaves in a similar fashion and, more generally, to compare 
its behaviour with the one obtained by cooling. To this aim we have computed 
$\chi$ for the three different lattice spacings adopted, {using configurations 
smoothed both with cooling and the gradient flow.} 
In~\Figref{chi_5.95} and \Figref{chi_6.2} the values of $\chi$ obtained for $\beta=5.95$ and $\beta=6.2$ are plotted 
against the deviation of the average plaquette from unity, 
which is proportional to the action density, 
\ie the variable that we have established as a ``thermometer'' 
to compare cooling and the gradient flow.

Apart from small deviations at the very beginning of the smoothing, 
the two determinations are completely equivalent: the agreement is perfect 
starting from $n_c = 8$ on the coarsest, and starting from $n_c = 3$ on
the finest explored lattice, see also Table~\ref{table2} for some representative numerical values.
On the other hand, such an agreement was already expected from the 
superposition of the two topological charge distributions,
shown in Fig.~\ref{qdistr}, since $\chi$ is just one of the moments
of this distribution.
On a larger $26^4$ lattice, the value of $a^4\chi\times 10^5$ for $n_c=21$ and 
$\beta=6.2$ is $1.983(17)$, in perfect agreement with the one reported in the table, 
obtained by using a $20^4$ lattice.

\begin{figure}[t!]
\includegraphics[width=0.9\columnwidth, clip]{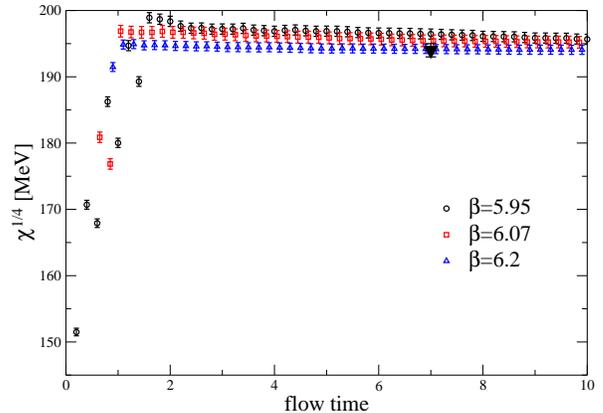}
\caption{Behaviour of the topological susceptibility under the gradient flow for the different lattice spacings adopted. 
The black filled triangle denotes a check for finite size effects performed on a $26^4$ lattice for the bare coupling 
value $\beta=6.2$.}
\label{chi_wilson}
\end{figure}

\begin{figure}
\includegraphics[width=0.9\columnwidth, clip]{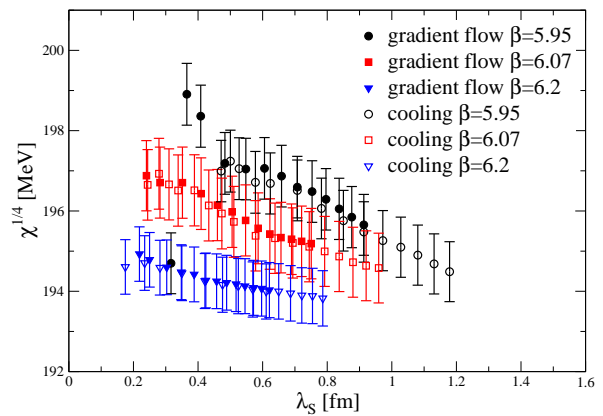}
\caption{Behaviour of the topological susceptibility under cooling and the gradient flow (zoom of the plateau region)
as a function of the smoothing length $\lambda_S$ defined by Eqs.\eqref{lambda_wilson}-\eqref{lambda_cooling}. 
}
\label{chi_lambda}
\end{figure}

A nice scaling of the topological susceptibility to the continuum limit is observed
in both cases (\ie both for cooling and the gradient flow), see Figs.~\ref{chi_wilson}-\ref{chi_lambda} 
for the case of the gradient flow, with extended plateaux around $\chi^{1/4} \sim 195$ MeV, even if 
an accurate estimate of the finite size and UV cutoff systematic
effects is not the purpose of this study.

\begin{table}
\begin{tabular}{|c|c|c|c|}
\hline $\beta$ & 5.95 & 6.07 & 6.2 \\ \hline\hline

$n_c=9$ &  10.91(17) &  4.644(83) &  1.998(30) \\ \hline
$\tau=3$ &  10.91(18) &  4.653(84) &  1.999(30) \\ \hline \hline

$n_c=21$ &  10.68(17) &  4.554(81) & 1.985(29) \\ \hline 
$\tau=7$  &  10.74(17) &  4.566(82) & 1.987(29) \\ \hline 
\end{tabular}
\caption{\label{table2}
Values of $a^4\chi\times 10^5$ for the three explored values of $\beta$ and 
for a couple of corresponding pairs of cooling steps and gradient flow time.
If no rounding is applied to define
$Q$, one still has perfect agreement between cooling and the gradient flow,
for instance for $\beta = 6.2$ one obtains  $a^4\chi \times 10^5 = 1.798(26)$ and  
$1.799(26)$ for $n_c = 21$ and $\tau = 7$, respectively.} 
\end{table}

\subsection{Comparison configuration by configuration}

We have shown that cooling and the gradient flow provide perfectly equivalent
results for average topological quantities, such as the topological
susceptibility. Here we want to make a more stringent test, comparing
the outcome of the gradient flow and of cooling configuration by configuration.

\begin{figure}[t!]
\includegraphics[width=0.9\columnwidth, clip]{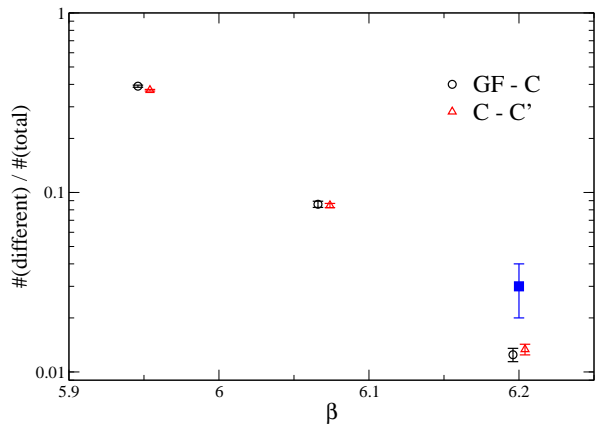}
\caption{Fraction of the configurations for which different results are obtained,
for the topological charge, using two different procedures. Circles refer to the 
comparison between cooling and gradient flow, while triangles refer to different cooling implementations 
(data points have been slightly shifted horizontally in order to distinguish them). The square dot
is the result of the comparison between cooling and the gradient flow on a $30^4$ lattice, \ie at fixed 
physical volume with respect to the $20^4$ lattice at $\beta=5.95$.
}
\label{wilson_c_c2_tris}
\end{figure}

First, we have determined the percentage of configurations where 
cooling and the gradient flow obtain different results for the global
topological content $Q$. This is reported in \Figref{wilson_c_c2_tris}:
the topological charges were estimated after $21$ cooling steps 
and after $7$ units of flow time, respectively, however results
are stable in a wide range of $n_c$.
The percentage is around 40\% on the coarsest lattice 
($\beta = 5.95$) and it rapidly decreases, seemingly exponentially in $\beta$,
reaching around 1\% on the finest lattice ($\beta=6.2$). 

One could suspect that this strong decrease is related to the variation 
of the physical volume; however, the effect of the volume change is in fact just a 
small contribution to this decrease. To check this point, we have performed simulations
on a $30^4$ lattice at bare coupling $\beta=6.2$ (which is approximately of the same 
physical size as the $20^4$ lattice with $\beta=5.95$) and we have found that the percentage 
of configurations on which cooling and the gradient flow do not agree in the determination of $Q$ 
is about $3\%$. Therefore, also at constant volume this quantity strongly decreases with the lattice spacing.
 
It is interesting to compare this rate of different 
determinations of $Q$
with the analogous rate obtained by comparing two 
slightly different versions of cooling, the one used in this 
paper and a simple variation, in which we just move 
the starting point of the cooling sweep from the origin to
the middle of the lattice.
Results are reported in \Figref{wilson_c_c2_tris} as well, and are 
completely equivalent with the previous ones, showing that the differences between cooling and 
the gradient flow are perfectly explainable in terms of the normal variations between different 
smoothing techniques, which take place when the starting configuration
presents some degree of coarseness and rapidly disappear as one approaches the continuum limit.

\begin{figure}[t!]
\includegraphics[width=0.9\columnwidth, clip]{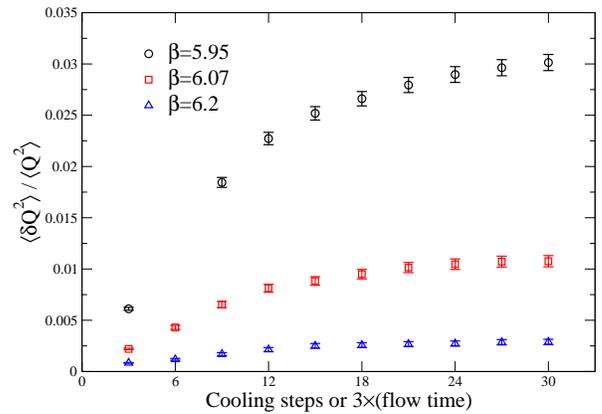}
\caption{Plot of $\langle \delta Q^2 \rangle / \langle Q^2 \rangle $ 
as a function of the number of cooling steps. 
See \Eqref{deltaQ2def} for definitions.}
\label{deltaQ2_on_Q2}
\end{figure}

\begin{figure}[t!]
\includegraphics[width=0.9\columnwidth, clip]{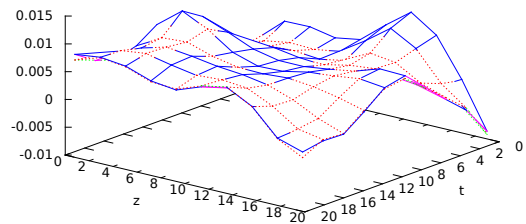}
\caption{Projection on the $z-t$ plane of the topological charge density for a $\beta=6.2$ 
configuration of total charge $Q=2$ after $21$ cooling steps (continuous line) or after $7$ 
units of gradient flow time (dotted line).}
\label{figura3d}
\end{figure}

A slightly different question is how much the different determinations 
of $Q$ between the two methods are relevant with respect to the 
global topological activity, taking place on a given lattice
at a given $\beta$ value.
In order answer this question, we have measured the quantity
\begin{equation}
\label{deltaQ2def}
 \langle \delta Q^2 \rangle \equiv \langle (Q^{(c)}-Q^{(gf)})^2 \rangle\, ,
\end{equation}
where $Q^{(c)}$ and $Q^{(gf)}$ are the corresponding (\ie at $n_c=3\tau$) estimates of the topological charge 
obtained by cooling and by the gradient flow, and we have normalized
it by the corresponding value of $\langle Q^2 \rangle$: in this way
we normalize with respect to possible variations of the topological 
activity due, for instance, to the different physical volumes. 
Numerical results for 
$\langle \delta Q^2 \rangle / \langle Q^2 \rangle$
are shown in \Figref{deltaQ2_on_Q2}: also in this case 
it is clear that the differences rapidly disappear 
as one approaches the continuum limit.

Finally, it is interesting to ask whether the observed 
agreement between cooling and the gradient flow is something that 
regards only the global topological charge of gauge configurations,
or whether the agreement holds true also at a local level.
The latter, of course, is a much stronger statement.
In \Figref{figura3d} we report the topological charge density,
projected on the $z-t$ plane, obtained on a typical configuration
where $Q^{(gf)} = Q^{(c)}=2$ (with $\beta = 6.2$, $n_c=21$ and
$\tau = n_c/3$). 
As one can appreciate, the agreement is also very good also at a local
level. Such a result may give hints that for observables 
not directly related to topology one could also obtain similar results
when adopting cooling or the gradient flow; however,
a systematic investigation of this possibility is left
to future studies.

\section{Discussion and conclusions}
\label{sec4}

The purpose of this study was to compare the gradient flow
and the discrete flow generated by 
standard cooling, with respect to the determination
of the topological properties of non-Abelian 
gauge theories on a lattice, with particular reference
to the case of the $SU(3)$ pure gauge theory.

To that aim, we have first established a relation
between the gradient flow time $\tau$ and the number of cooling
steps $n_c$, so that the plaquette action density, which is the quantity
minimized by both flows, coincides.  
The relation $\tau \simeq n_c/3$ holds within a good precision
and already after a few cooling steps; such a relation is also in 
agreement with a perturbative estimate, which is expected to be 
valid in the limit of smooth fields and to depend on the
dimensionality of the system ($\tau = n_c/(D - 1)$ 
where $D$ is the number of space-time dimensions), but not on the 
details of the gauge group.

We have proven that, after a very few transient 
cooling steps, the two flows lead to equivalent results,
the transient region rapidly decreasing as the continuum
limit is approached.
This assertion is true at various degrees of strength.
It is true for average quantities: we have checked 
this assertion with some accuracy for the topological
susceptibility; however, given the superposition of the two
probability distributions (see \Figref{qdistr}), we expect
this to be true also for the higher order moments,
which are needed to specify the $\theta$ dependence of the 
theory~\cite{b4-1,b4-2,b4-3,b4-4, vicari_rep}

This expectation is also supported by the fact that, 
at a stronger level, even the discrepancies 
in the determination of $Q$, which are found			
configuration by configuration, rapidly become irrelevant
as the continuum limit is approached, and already for 
$a \sim 0.1$~fm. Moreover, the local profiles of topological
charge densities, obtained 
by the two smoothing methods
on sample configurations, 
are also very close to each other.

It is important, at this point, to stress that these conclusions do not depend
on the specific prescriptions adopted to define the lattice 
topological charge or to perform the continuum limit.
The outcome of this study is that, at every fixed lattice spacing, cooling and gradient flow give
the same result provided the number of cooling steps $n_c$ and the flow time $\tau$ are related 
by $\tau=n_c/3$ (or, equivalently, the smoothing cutoff $\lambda_S$ is chosen according to
\Eqref{lambda_wilson} and \Eqref{lambda_cooling}).

Given the equivalence of the two
procedures from a practical point of view, and at least
regarding topological quantities, the choice of the method 
to be used in future simulations relies on the computational efficiency.

While for some applications the computational cost of
both cooling and the gradient flow is negligible (like, \eg, for scale setting by the 
$t_0$ parameter \cite{Luscher_wf1}) there are cases in which this is not
true. As a typical example we mention the evaluation of the higher momenta
of the topological charge and, in particular, the computation of the renormalization 
group invariants commonly denoted by $b_{2n}$ (see, \eg, \cite{vicari_rep}), whose determination 
requires $\mathcal{O}(10^5\div10^6)$ independent determinations of $Q$. 
This makes even the pure gauge simulations far from trivial and the computational 
efficiency of the method used to estimate the topological charge becomes a crucial ingredient.

In particular, using the established relation $\tau \simeq n_c / 3$,
we can compare the execution time of three cooling steps with the time needed to perform an unity of 
gradient flow time evolution, obtaining
${\mathrm{cpu\_time}(\tau=1)}/{\mathrm{cpu\_time}(n_c=3)} \simeq 130$.
Clearly these estimates depend on the specific
integrator adopted for the gradient flow and, in particular, 
adaptive integrators make it possible 
to obtain an $\mathcal{O}(10)$ 
speedup with respect to the third order RK solver (see Refs.~\cite{FR1, FR2}). 
Nevertheless, cooling remains about one order 
of magnitude cheaper than the gradient flow.

Of course, one should consider that the gradient flow has 
advantages with respect to techniques like cooling, related
to the fact that it has an 
associated differential equation, 
which clearly appear whenever an analytical
treatment of the smoothing process is required,
like, for instance, in the analysis of the renormalization
properties of the smoothed fields~\cite{Luscher:2011bx}.
Moreover, the gradient flow can be consistently extended
to the presence of dynamical fermion fields~\cite{Luscher:2013cpa}.
We refer to~\cite{Luscher:2013vga} for a recent review of 
present and future perspectives of the gradient flow.

Finally, given the agreement of the topological 
charge density also at a local level, in the future
one should better investigate the relation between
the two smoothing procedures for other physical quantities as well.

\section*{Acknowledgments}

We thank A.~Di Giacomo, M.~Mesiti, F.~Negro, F.~Sanfilippo and E.~Vicari for useful discussions.
Numerical simulations have been performed on 
the CSNIV cluster at the Scientific Computing Center at INFN-Pisa.

\appendix

\section{A simple diffusive model}
\label{appendA}

Let us consider a massless real scalar field, $\phi(n)$, on
a three dimensional isotropic 
cubic lattice,
where $n \equiv (n_x,n_y,n_z)$,  
and with the associated action
\begin{equation*}
S = \sum_{n,\hat j} \frac{1}{2} \big[\phi(n + \hat j) - \phi(n)\big]^2
\end{equation*}
where $\hat j$ runs over the three positive directions
and $n + \hat j$ indicates, as usual, the lattice site 
which is the nearest neighbor of $n$ in the forward $\hat j$ direction.
We will now consider the gradient flow for such a theory,
and the differential equation obtained by it in the limit
of smoothly varying fields. Then we will do the same
in the case of cooling, and compare the two differential 
equations.

The gradient flow is defined by adding a dependence 
of $\phi$ on a fictitious time $\tau$ and letting
\begin{eqnarray}\label{model2}
\frac{\partial \phi(n,\tau)}{\partial \tau} &=& 
- \frac{\partial S(\tau)}{\partial \phi(n,\tau)} 
\\
&=& 
\sum_{\hat j} \big[\phi(n + \hat j,\tau) + \phi(n - \hat j,\tau)\big] 
- 6 \phi(n , \tau) \, . \nonumber
\end{eqnarray}
In the limit of smoothly varying fields, we can 
take a continuum description and, letting $x_j \equiv a\, n_j$,
where $a$ is the lattice spacing, 
change the notation $\phi (n,\tau) \to \phi(x,\tau)$. The field
on nearest neighbor sites can 
be Taylor expanded
\begin{equation*}
\phi(n + \hat j,\tau) \simeq \phi (x,\tau) + 
a \frac{\partial \phi}{\partial x_j} + \frac{a^2}{2} 
\frac{\partial^2 \phi}{\partial x_j^2} 
\end{equation*}
so that the flow equation, Eq.~(\ref{model2}), takes the simple 
form of a diffusive (heat) equation:
\begin{equation}\label{model4}
\frac{\partial \phi(x,\tau)}{\partial \tau} 
\simeq a^2 \nabla^2 \phi(x,\tau) 
\end{equation}
where $\nabla^2$ is the 3D Laplacian operator.

Let us now consider cooling, in which the field
is evolved by local minimization steps, which are
iterated by sweeping all lattice sites at each
cooling step. Let us call
$\phi(n,n_c)$ the field obtained after $n_c$ steps.
If the lattice sites are visited in the positive lexicographic order,
then it is easy to verify that the cooling equation is
\begin{equation}\label{model5}
\phi(n,n_c + 1) = \frac{1}{6} \sum_j
\big[\phi(n + \hat j,n_c) + \phi(n - \hat j,n_c + 1)\big] \, 
\end{equation}
where we have taken into account that part of the nearest neighbor sites
have already undergone the cooling step under consideration.
In order to write a corresponding differential equation, we now consider
that, in the limit of smoothly varying fields, the evolution
generated by cooling is also smooth, so we can Taylor expand
in terms of a cooling timeas well, defined by 
$\tau_c \equiv a_\tau n_c$, where $a_\tau$ is a 
fictitious temporal spacing that will be eventually set to 1.
We can therefore substitute
\begin{eqnarray*}
\phi(n,n_c + 1) &\simeq&
\phi(x,\tau_c) + 
a_\tau \frac{\partial \phi}{\partial \tau_c}\,  \\
\phi(n-\hat j,n_c + 1) &\simeq&
\phi(x,\tau_c) + 
a_\tau \frac{\partial \phi}{\partial \tau_c} +
a \frac{\partial \phi}{\partial x_j} + \frac{a^2}{2} 
\frac{\partial^2 \phi}{\partial x_j^2}\, \\
\phi(n + \hat j,n_c) &\simeq&
\phi(x,\tau_c) + 
a \frac{\partial \phi}{\partial x_j} + \frac{a^2}{2} 
\frac{\partial^2 \phi}{\partial x_j^2} \; .
\end{eqnarray*}
We notice that, since we are dealing with a diffusion process,
in which spatial distances scale like the square root of the
diffusion time, it is consistent to Taylor expand 
at the linear order in time and at the quadratic order
in spatial derivatives.
Putting everything together, Eq.~(\ref{model5}) becomes
\begin{equation}\label{model7}
a_\tau \frac{\partial \phi(x,\tau_c)}{\partial \tau_c}  \simeq
 \frac{1}{3} a^2 \nabla^2 \phi(x,\tau)
\end{equation}
which after setting $a_\tau = 1$ teaches us that the relation between
the cooling time $\tau_c$ and the gradient flow time $\tau$ is 
\begin{equation*}
\tau_c = n_c \simeq 3 \tau \, , 
\end{equation*}
meaning that 3 cooling steps correspond to 1 unit of gradient flow time.
It should be clear that the factor 3 would have been a factor 6,
had we not taken into account the additional cooling time 
dependence of half of the nearest neighbor fields.

\end{document}